\documentstyle[12pt]{article}

\begin{document}

\begin{titlepage}
\begin{center}
        {\large \bf Dirac  spinor in a nonstationary G\"odel-type
        cosmological universe}

\vspace{1.0 cm}

V\'{\i}ctor M. Villalba\\
Centro de F\'{\i}sica\\
Instituto Venezolano de Investigaciones Cient\'{\i}ficas, IVIC\\
Apdo 21827, Caracas 1020-A, Venezuela

\vspace{1.0 cm}
\end{center}
\begin{abstract}
In the present article we solve, via separation of variables, the
massless Dirac
equation in a nonstationary rotating, causal G\"odel-type cosmological
universe,
having a constant rotational speed in all the points of the space. We
compute
the frequency spectrum. We show that the spectrum of massless Dirac
particles is
discrete and unbounded.
\end{abstract}
\vspace{1.0cm}
\end{titlepage}

During the last years there has been a growing interest in study of
quantum
effects associated with scalar and spinor particles in cosmological
universes. In this direction a particular attention has been devoted to
the
analysis of homogeneous universes as well as of rotating space-times.

Obviously, in order to analyze quantum effects on curved backgrounds is
necessary to carry out a careful study of the one-particle states, that
is,
a detailed investigation of the exact solution of the relativistic wave
equations in curved space-times.

Despite the considerable effort dedicated  to the search of exact
solutions
of relativistic wave equations in curved space-times, only a few
examples
are available in the literature where the solutions  are expressed in
closed
form, being the background fields where this one is possible associated
with
diagonal metrics \cite{victor3} or in the best ot the cases with the
type D
Petrov spaces\cite{kamran}. A different family of universes, where
scalar
particles as well as Dirac particles have been analyzed, is the
so-called
G\"odel-type space-times. The interest in this kind of models is related
to
the study of the possible  effects of the  rotation in the formulation
of a
quantum field theory on curved spaces. In this framework some exact
solutions of the massless Dirac equation
\begin{equation}
\label{wey}\gamma ^\alpha \nabla _\alpha \Psi =0,
\end{equation}
\noindent with the chirality condition
\begin{equation}
\label{chi}(1-i\gamma _5)\Psi =0,
\end{equation}
have been obtained. Regretfully, due to the structure of the system of
equations to be decoupled and solved, no exact solutions in terms of
special
functions are available for the massive case. Among the space-times
where (%
\ref{wey}) and (\ref{chi}) have been solved we have  the  G\"odel
universe
\cite{Pimentel1}
\begin{equation}
ds^2=-(dt+e^{ar}d\theta )^2+dr^2+\frac 12(e^{ar}d\theta )^2+dz^2,
\end{equation}
\noindent also the universe associated with The Einstein-Maxwell
equations
in the presence of a perfect fluid and a sinusoidal electromagnetic
field
with vanishing rest charge density \cite{victor1}

\begin{equation}
ds^2=-(dt+axdy)^2+dx^2+dy^2+dz^2,
\end{equation}
other metrics where the Klein-Gordon and the Weyl equation are also
soluble
in terms of hypergeometric functions are\cite{Krori}

\noindent a) The Som-Raychaudhuri metric
\begin{equation}
ds^2=-dt^2+dr^2+dz^2-2r^2d\phi dt+r^2(1-r^2)d\phi ^2,
\end{equation}

\noindent b) The Hoenselaers-Vishveshwara metric
\begin{equation}
ds^2-dt^2+dr^2+dz^2-\frac 12(c-1)(c-3)d\phi ^2-2(c-1)d\phi dt,
\end{equation}

\noindent c) The Rebou{\c c}as metric
\begin{equation}
ds^2=-dt^2+dr^2+dz^2+4\cosh 2rd\phi dt\ -(3\cosh {}^22r+1)d\phi ^2,
\end{equation}
in all the above cases the scalar wave as well as the Dirac spinor
present
an oscillating behavior in the radial variable.

Recently \cite{Panov1} the massless Klein-Gordon equation.
\begin{equation}
\label{uno}\nabla ^\alpha \nabla _\alpha \Phi =0,
\end{equation}
has been studied in a cosmological rotating model associated with the
metric,

\begin{equation}
\label{one}ds^2=-dt^2+c^2t^2(dx^2+\lambda
e^{2mx}dy^2+dz^2)+2cte^{mx}dydt,
\end{equation}
where c, $\lambda ,$ and $m$ are positive constants The rotational speed
of
the model given by eq. (\ref{one}) is
\begin{equation}
\omega =\frac m{2ct\sqrt{\lambda +1}}.
\end{equation}
Among the advantages of the metric (\ref{one}) we have that the model is
causal , i.e., there are not closed timelike lines. Exact solutions of
eq. (%
\ref{uno}), and the corresponding frequency spectrum analysis for the
metric
(\ref{one}) with $\lambda =0,$ have been presented by
Panov\cite{Panov2}. It
is the purpose of the present article to obtain exact solutions of the
massless Dirac equation in the background field given by eq. (\ref{one})
(even for $\lambda\neq 0$) and to compute the corresponding frequency
spectrum.

The covariant generalization of the massless Dirac equation in curved
space-time is,
\begin{equation}
\label{ecua}\left[ \gamma ^\mu (\partial _\mu -\Gamma _\mu )\right] \Psi
=0,
\end{equation}
where the curved Dirac matrices satisfy the anticommutation relations
\begin{equation}
\left\{ \gamma ^\mu ,\gamma ^\nu \right\} _{+}=2g^{\mu \nu },
\end{equation}
and the spinor connection $\Gamma _\lambda $ is\cite{Brill}

\begin{equation}
\label{conex}\Gamma _\lambda =\frac 14g_{\mu \alpha }\left( \partial
_\lambda h_\nu ^{\ k}h_{\ k}^\alpha -\Gamma _{\nu \lambda }^\alpha
\right)
s^{\mu \nu }.
\end{equation}
In order to compute (\ref{conex}) we have to choose a particular tetrad
associated with the line element given by (\ref{one}). A suitable
election
is,
\begin{equation}
\label{tetrad}\gamma ^0=\tilde \gamma ^0+\frac 1{\sqrt{\lambda
+1}}\tilde
\gamma ^2,\ \gamma ^1=\frac 1{ct}\tilde \gamma ^1,\ \gamma
^2=\frac{e^{-mx}}{%
ct\sqrt{\lambda +1}}\tilde \gamma ^2,\ \tilde \gamma ^3=\frac
1{ct}\tilde
\gamma ^3,
\end{equation}
after substituting (\ref{tetrad}) into (\ref{conex}), we have that the
Dirac
equation takes reads,
\begin{equation}
\label{Dirac}\left\{ \tilde \gamma ^0+\frac 1{\sqrt{\lambda +1}}\tilde
\gamma ^2)\partial _t+\frac 1{ct}\tilde \gamma ^1\partial
_x+\frac{e^{-mx}}{%
ct\sqrt{\lambda +1}}\tilde \gamma ^2\partial _y+\frac 1{ct}\tilde \gamma
^3\partial _z\right\} \Phi =0,
\end{equation}
where we have introduced the spinor $\Phi $ given by the expression
\begin{equation}
te^{mx/2}\Psi =\Phi .
\end{equation}
After introducing the new time variable $\eta $ which is related to $t$
by
the expression
\begin{equation}
t=\exp (c\eta ),
\end{equation}
we have that (\ref{Dirac}) takes the form
\begin{equation}
\label{operator}\left\{ -\omega \tilde \gamma ^0-i\tilde \gamma
^1\partial
_x+\frac 1{\sqrt{\lambda +1}}(\frac{k_y}{e^{mx}}-\omega )\tilde \gamma
^2+k_z\tilde \gamma ^3\right\} \Phi _0=0,
\end{equation}
where the spinor $\Phi _0$ is related to $\Phi $ as follows,
\begin{equation}
\Phi =\Phi _0e^{i(k_yy+k_zz-\omega t)},
\end{equation}
the operator (\ref{operator}) can be written as a sum of two commuting
operators $\hat K_1$ and $\hat K_2$ given by the
expressions\cite{victor2}
\begin{equation}
\label{k1}\hat K_1=\left( -\omega \tilde \gamma ^0+k_z\tilde \gamma
^3\right) \tilde \gamma ^0\tilde \gamma ^3,
\end{equation}
\begin{equation}
\label{k2}\hat K_2=\left( -i\tilde \gamma ^1\partial _x+\frac 1{\sqrt{%
\lambda +1}}(\frac{k_y}{e^{mx}}-\omega )\tilde \gamma ^2\right) \tilde
\gamma ^0\tilde \gamma ^3,
\end{equation}
where $\hat K_1$ and $\hat K_2$ act on the spinor $\Xi $ as follows,
\begin{equation}
\label{comm}\hat K_1\Xi =-\hat K_2\Xi =i\kappa \Xi ,\ \ \Xi =\tilde
\gamma
^0\tilde \gamma ^3\Phi _0.
\end{equation}
Choosing to work in the Jauch and Rohrlich\cite{Jauch} Dirac matrices'
representation,
\begin{equation}
\label{repr}\tilde \gamma ^0=\left(
\begin{array}{cc}
-i & 0 \\
0 & i
\end{array}
\right) ,\ \tilde \gamma ^i=\left(
\begin{array}{cc}
0 & \sigma ^i \\
\sigma ^i & 0
\end{array}
\right) ,\ i=1,2,3
\end{equation}
we have that, taking into account (\ref{k1}) and (\ref{repr}), the
spinor $%
\Xi $ takes the following block structure.
\begin{equation}
\label{block}\Xi =\left(
\begin{array}{c}
\Xi _1 \\
\Xi _2
\end{array}
\right) =\left(
\begin{array}{c}
\Xi _1 \\
\frac{-i\omega }{k_x+\kappa }\sigma _3\Xi _1
\end{array}
\right) ,\quad \Xi _1=\left(
\begin{array}{c}
\alpha  \\
\beta
\end{array}
\right)
\end{equation}
where the constant of separation appearing in (\ref{comm}) satisfies the
relation,
\begin{equation}
\kappa ^2=k_z^2-\omega ^2,
\end{equation}
after substituting (\ref{block}) into $\hat K_1\Xi =i\kappa \Xi $ with
$\hat
K_1$ given by (\ref{k2}) we obtain the following system of coupled
differential equations,
\begin{equation}
\label{odin}\left[ \frac d{dx}+\frac 1{\sqrt{\lambda
+1}}(\frac{k_y}{e^{mx}}%
-\omega )\right] \beta -i\kappa \alpha =0,
\end{equation}
\begin{equation}
\label{dva}\left[ \frac d{dx}-\frac 1{\sqrt{\lambda
+1}}(\frac{k_y}{e^{mx}}%
-\omega )\right] \alpha +\ i\kappa \beta =0,
\end{equation}
then, the spinor solution $\Psi _c$ of the massless Dirac equation with
the
chirality condition (\ref{chi}) can be written as follows,
\begin{equation}
\Psi _c=e^{-(c\eta +mx/2)}(1+i\gamma _5)\left(
\begin{array}{c}
-\frac \omega {k_x+\kappa }\Xi  \\
i\sigma ^3\Xi
\end{array}
\right)
\end{equation}
the solution of (\ref{odin}), (\ref{dva}) can be expressed in terms of
confluent hypergeometric functions $M(a,b,z)$,
\begin{equation}
\alpha =-ia_0\frac{k_z-\omega }{m\sqrt{\lambda +1}}e^{-k_y\eta /m\sqrt{%
\lambda +1}}\eta ^{k_z/m\sqrt{\lambda +1}}M(-\frac{\omega +k_z}{m\sqrt{%
\lambda +1}}+1,\frac{2k_z}{m\sqrt{\lambda +1}}+1,\frac{2k_y\eta
}{m\sqrt{%
\lambda +1}})
\end{equation}
\begin{equation}
\beta =a_0e^{-k_y\eta /m\sqrt{\lambda +1}}\eta ^{k_z/m\sqrt{\lambda
+1}}M(-
\frac{\omega +k_z}{m\sqrt{\lambda +1}},\frac{2k_z}{m\sqrt{\lambda
+1}}+1,
\frac{2k_y\eta }{m\sqrt{\lambda +1}})
\end{equation}
in order to have regular solutions of $\alpha $ and $\beta $ for any
value
of $\eta ,$ we must impose the condition
\begin{equation}
\omega +k_z=(n+1)m\sqrt{\lambda +1},
\end{equation}
which determines the frequency spectrum for massless Dirac particles in
the
background field (\ref{one}). We have to notice that the frequency
spectrum
is discrete when $k_z=0$

A cosmological model closed related to that one described by (\ref{one})
is $%
.$
\begin{equation}
\label{three}ds^2=R^2(-d\tau ^2+dx^2+dz^2+2e^{mx}dyd\tau ),
\end{equation}
where $R=R(\tau )$ is an arbitrary differentiable positive function, and
the
constant $m$ is different from zero. In the present case there is a null
coordinate, and also an expansion factor $R(\tau )$ is present. In order
to
study the massless Dirac equation in the metric (\ref{three}) we are
going
to work in the following tetrad gauge,
\begin{equation}
\label{tetrada}\gamma ^0=\frac 1R\left( \tilde \gamma ^0+\tilde \gamma
^2\right) ,\ \gamma ^1=\frac 1R\tilde \gamma ^1,\ \gamma
^2=\frac{e^{-mx}}%
R\tilde \gamma ^2,\ \tilde \gamma ^3=\frac 1R\tilde \gamma ^3.
\end{equation}
After computing the spinor connections and substituting them into
(\ref{ecua}%
) we arrive at,
\begin{equation}
\label{operators}\left\{ -\omega \tilde \gamma ^0-i\tilde \gamma
^1\partial
_x+(\frac{k_y}{e^{mx}}-\omega )\tilde \gamma ^2+k_z\tilde \gamma
^3\right\}
\Phi _1=0,
\end{equation}
where $\Phi _1$ is,
\begin{equation}
\Phi _1=e^{\frac{mx}2}e^{\frac 34R^2}\Psi .
\end{equation}
Noticing that (\ref{operator}) reduces to (\ref{operators}) when we put
$%
\lambda =0$ we have that in the present case the frequency spectrum
takes
the form
\begin{equation}
\label{spec}\omega +k_z=(n+1)m.
\end{equation}

\end{document}